\documentclass[aps,prb,twocolumn,showpacs,amsmath,amssymb]{revtex4}
\usepackage{graphicx}
\usepackage{dcolumn}
\usepackage{bm}

\begin{document}

\preprint{APS/123-QED}

\title{Study of Field-Induced Magnetic Order in Singlet-Ground-State Magnet CsFeCl$_3$}

\author{Mitsuru Toda$^1$}
\author{Yutaka Fujii$^2$}
\author{Shinji Kawano$^3$}
\author{Takao Goto$^4$}
\author{Meiro Chiba$^2$}
\author{Shizumasa Ueda$^5$}
\author{Kenji Nakajima$^6$}
\author{Kazuhisa Kakurai$^7$}
\author{Jens Klenke$^8$}
\author{Ralf Feyerherm$^8$}
\author{Matthias Meschke$^8$}
\author{Hans Anton Graf$^8$}
\author{Michael Steiner$^8$}

\affiliation{%
$^1$Institute of Materials Structure Science, KEK, Tsukuba, Ibaraki 305-0801, Japan\\
$^2$Department of Applied Physics, Fukui University, Fukui 910-8507, Japan\\
$^3$Research Reactor Institute, Kyoto University, Kumatori 590-0494, Japan\\ 
$^4$Graduate School of Human and Environmental Studies, Kyoto University, Kyoto 606-8501, Japan\\
$^5$Institute of Advanced Energy, Kyoto University, Uji, Kyoto 611-0011, Japan\\ 
$^6$Center for Neutron Science, Japan Atomic Energy Research Institute, Tokai, Ibaraki 319-1195, Japan\\ 
$^7$Advanced Science Research Center, Japan Atomic Energy Research Institute, Tokai, Ibaraki 319-1195, Japan\\ 
$^8$BENSC, Hahn-Meitner-Institut Berlin, D-14109 Berlin, Germany
}

\date{\today}
\begin{abstract}
The field-induced magnetic order in the singlet-ground-state system CsFeCl$_3$ has been studied by measuring magnetization and neutron diffraction. The field dependence of intensity for the neutron magnetic reflection has clearly demonstrated that the field-induced ordered phase is described by the order parameter $\langle S_x \rangle$. A condensate growth of magnons is investigated through the temperature dependence of $M_z$ and $M_{\perp}$, and this ordering is discussed in the context of a magnon Bose-Einstein condensation. Development of the coherent state and the static correlation length has been observed in the incommensurate phase in the field region of $5 < H < 6$ T at 1.8 K. At $H > H_{\rm c}$, a satellite peak was found in coexistence with the commensurate peak at the phase boundary around 10 T, which indicates that the tilt of the $c$-axis would be less than $\sim 0.5^{\circ}$ in the whole experiments.

\end{abstract}

\maketitle

\section{\label{sec:level1}INTRODUCTION}
 Recently, many researchers have revised the field-induced magnetic order in the singlet-ground-state system with great interests. Especially, the field-induced magnetic order in the dimer spin system have been impressively investigated in the context of a Bose-Einstein condensation (BEC) of magnetic excitations (magnons)\cite{nikuni}. A well-known singlet-ground-state system in the hexagonal compound CsFeCl$_3$ is constituted by the ground state $|S_z=0\rangle$ and the first excited states $|S_z=\pm1\rangle$ of a fictitious spin $S=1$ of an Fe$^{2+}$ ion, each state being separated by single-ion anisotropy energy $D$, owing to a trigonal symmetrical distortion of Cl$^-$ ions along the $c$-axis ($c {\small /\!\!/} z$). Due to the ferromagnetic exchange interaction ($J_1/k_{\rm B} \approx 4$ K) along the $c$-axis, which is relatively larger than the antiferromagnetic one in the $c$-plane ($J_2/k_{\rm B} \approx -0.2$ K, triangular lattice antiferromagnet), the spin system becomes quasi-one dimensional. The effect of $D$ is predominant over that of exchange interaction energies, thus the ground state is nonmagnetic down to zero K. However, the magnetic phase transition can be induced by applying the external magnetic field along the $c$-axis, then the three dimensional long-range ordered (3D-LRO) state appears around the level crossing field ($H_{\rm c}=7.5$ T) of $|S_z=0\rangle$ and $|S_z=+1\rangle$ states at $T_{\rm N} \sim 2$ K, in the field region of $4 \le H \le 11$ T\cite{haseda}. The theoretical study\cite{tsuneto} has shown that the phase transition is considered to be of the second order and the ordered state is characterized by the non-vanishing value of $\langle S_x \rangle$; and also, similarity of the present case to the field-induced order in a dimer model was pointed out.

 Magnetic excitations and the magnetic phase transition in CsFeCl$_3$ have been extensively investigated experimentally by the neutron scattering and NMR experiments so far\cite{yoshizawa,steiner,schmid,knop,chiba1,toda1,toda2}. The excitation energy decreases with lowering temperature, and development of the short range order has been found at the wave vector $\bm{q} \approx $(1/3 1/3 0) (${\it K}$-point)\cite{yoshizawa}. Suzuki described these magnetic behaviors by the dynamical correlated effective field approximation (DCEFA) theory, which takes into account the effect of a fluctuation and correlation via the parameter $\alpha$\cite{suzuki}. The nuclear magnetic relaxation time $T_1$ of $^{133}$Cs has been measured, and the relaxation mechanism through the magnetic excitation has been clarified\cite{chiba1,toda1,toda2}. Comparing the temperature dependence of $T_1$ in the field region below and above 4 T, it is suggested that the large numbers of magnetic excitations are created and a life time of these excitations becomes short in the critical region at $H > 4$ T\cite{toda1,toda2}. Magnetization measurements ($H {\small /\!\!/} c$) have been performed up to 33 T\cite{hori,chiba1}, and especially the metamagnetism has been paid particular attention.
 
 As for the 3D-LRO state, an incommensurate-commensurate phase transition has been already studied by neutron diffraction up to 5.5 T\cite{knop}($H {\small /\!\!/} c$); however, the research has been limited in the vicinity of the phase boundary, owing to the performance limitations of cryostats. 
In this paper, we have focused on the magnetic behavior in the 3D-LRO state, and report the neutron diffraction and magnetization studies under the magnetic field up to 10 T which completely exceeds $H_{\rm c}$.
 
\section{\label{sec:level2}EXPERIMENT}
  Samples have been prepared in the laboratory at the Institute of Advanced Energy (IAE), Kyoto University. The equimolar 99.9$\%$ CsCl and FeCl$_2$ powder samples were mixed in the ampoule; and after dehydrated using HCl gas\cite{hirakawa}, single crystals of a volume of 2$\sim$3 cm$^3$ large were separated out by the Bridgman method. Clear hexagonal cleavage planes were seen, and the color of the small crystal ($\sim$0.2 cm$^3$) was transparent dark green. The mosaic width was about $\sim 0.5 ^{\circ}$, and the hexagonal lattice parameters as used is $a$=7.237\r{A} (the lattice parameter of $c$-axis is not obtained) in the neutron experiments. 
  
  The magnetization measurements have been performed using a 9 T SQUID magnetometer at Berlin Technical University, which was built on the basis of standard construction principles, but supplied with additional features to allow measurements up to 9 T. The measured sample was the needle shaped crystal of samplemass 0.61 mg. The field dependent signal resolution is in the range of $5 \times 10^{-5}$ emu.

 The neutron diffraction experiments were carried out on the triple-axis spectrometer E1 at the Berlin Neutron Scattering Centre (BENSC), Hahn Meitner Institute (HMI), Berlin. The wavelength $\lambda$=2.4153\r{A} was selected by Bragg reflection at a pyrolytic graphite monochromator for the incoming beam, and the analyser was adjusted to measure the elastic part of the scattering. The used collimation was 70$^{\prime}$-80$^{\prime}$-60$^{\prime}$-60$^{\prime}$. The sample was mounted in the new high-field superconducting magnet VM1 built by Oxford Instruments, which with a maximum field of 17 T is actually the strongest steady-state magnet available for neutron scattering measurement. The preliminary experiments have been carried out on the triple-axis spectrometer PONTA at Institute for Solid State Physics (ISSP), Japan Atomic Energy Research Institute (JAERI), Tokai.
 
\section{\label{sec:level3}RESULTS AND DISCUSSION}
\subsection{\label{subsec:level1}Magnetization}  
\begin{figure}[htb]
\begin{center}
\includegraphics[width=8cm]{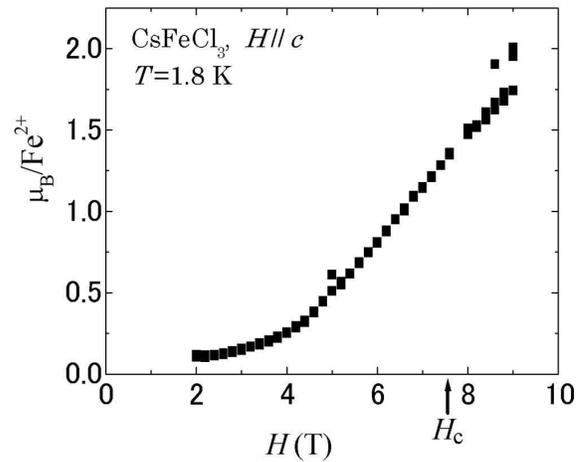}
\end{center}
\caption{Magnetization $M_z$ data as a function of external magnetic field ($H {\small /\!\!/} c$) at 1.8 K. Dissipation of the data at $H \approx 9$ T might be the experimental error.}
\label{f0}
\end{figure}

\begin{figure}[htbp]
\begin{center}
\includegraphics[width=8cm]{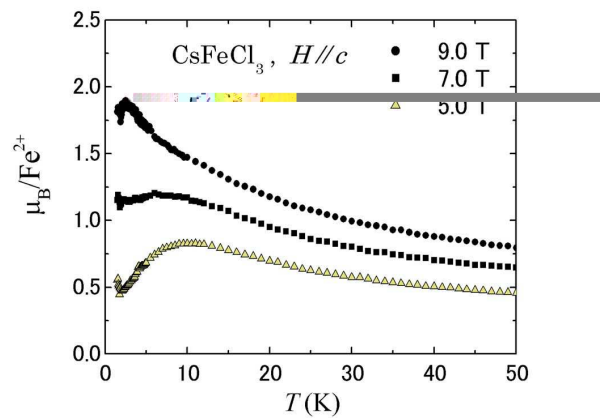}
\end{center}
\caption{Magnetization $M_z$ data as a function of temperature at 5, 7 and 9 T ($H {\small /\!\!/} c$).}
\label{f4}
\end{figure}

 First we report the results of the magnetization measurements. Figure \ref{f0} shows the data at $T=1.8$ K, plotted as a function of field ($H {\small /\!\!/} c$). 
At such low temperature, most of spins should be distributed at the ground state; therefore, $M_z$ is nearly equal to zero at low fields. Then, $M_z$ shows explicit increase above 4 T in the ordered phase, and is approximately linear up to 9 T. The value of magnetization is $\sim$ 1.3 $\mu_{\rm B}$ at $H_{\rm c}$, which is approximately half the magnetic moment expected from $g^{/\!\!/}=2.54$ \cite{suzuki}.  
 Figure \ref{f4} shows the results of the magnetization measurements as a function of temperature at 5, 7 and 9 T. 
At 5 T, the broad peak is seen at $T \approx 10$ K; this is mainly due to the thermal activation type term of $\exp(-E_1/k_{\rm B}T)$, where $E_1$ is the gap energy between the ground state $|S_z=0\rangle$ and the first excited state $|S_z=+1\rangle$. This broad peak clearly diminishes at 9 T, reflecting the change of the ground state from $|S_z=0\rangle$ to $|S_z=+1\rangle$.  Furthermore, the magnetization curves show bends and the signs of the derivatives change at 5 and 9 T  at $T \approx T_{\rm N}$.

 The clear temperature variations of $M_z$ are seen in the 3D-LRO state; however, when each spin is regarded as being independent in the molecular field approximation (MFA), the magnetization does not depend on temperature\cite{tsuneto}. The same feature has been found in the dimer spin system TlCuCl$_3$\cite{nikuni}. We expect that the temperature variations of $M_z$ result from the many body effect. Considering that the magnetic order is derived from the softening of magentic excitations, the coherent state of $|S_z=0\rangle$ and $|S_z=+1\rangle$ will develop following the condensation growth of zero energy magnetic excitations. Then $\langle S_z \rangle$ will increase with decreasing temperature in the ordered phase by the contribution of $|S_z=+1\rangle$ component.

\subsection{\label{subsec:level2}Neutron Diffraction} 
 Next we show the results of the high field neutron diffraction experiments. The previous NMR or neutron experiments resulted in failure frequently, since the large magnetic field inclines the misalignment of the $c$-axis furthermore. In this experiments, the single crystal of a volume of $\sim$1 cm$^3$ was attached to the sample holder, and combined into the micro-goniomator (Kohzu-Seiki co.); then the obliquity of the $c$-axis was controlled below $\sim 0.3 ^{\circ}$ at zero field. This configuration was very important to observe the incommensurate state\cite{toda3}. The former NMR experiments\cite{chiba2} have shown that the ordered phase region reduces in the phase diagram and moves to the lower fields while increasing the obliquity up to $10 ^{\circ}$. 
  
\begin{figure}[htbp]
\begin{center}
\includegraphics[width=8cm]{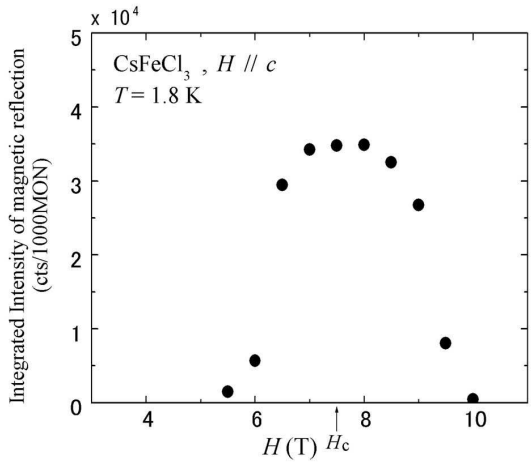}
\end{center}
\caption{The dependence of integrated intensity on external magnetic field ($H {\small /\!\!/} c$) for the neutron magnetic reflection at $T$=1.8 K.}
\label{f1}
\end{figure}

\begin{figure}[htbp]
\begin{center}
\includegraphics[width=8cm]{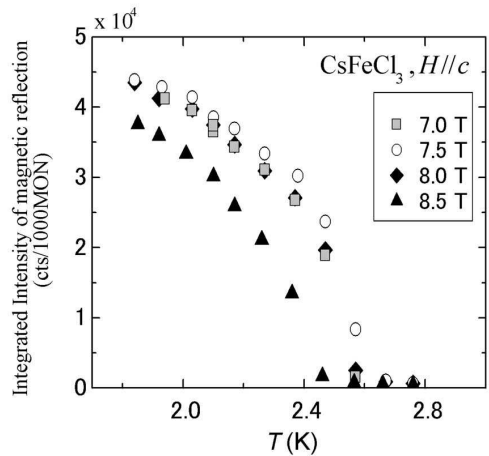}
\end{center}
\caption{ The dependence of integrated intensity on temperature for the neutron magnetic reflection at 7.0, 7.5, 8.0 and 8.5 T ($H {\small /\!\!/} c$).}
\label{f5}
\end{figure}

 Figure \ref{f1} shows the dependence of integrated intensity on external magnetic field ($H {\small /\!\!/} c$) for the neutron magnetic reflection at $T$=1.8 K.
 We have always confirmed the incommensurate state at $H \approx$ 5 T, before increasing the magnetic field to induce the commensurate state. The intensity of magnetic (1/3 1/3 0) reflection, which is directly related to the amplitude of $\langle S_x \rangle$, drastically increased and decreased almost symmetrically with respect to $H_{\rm c}$. Comparison with the phase diagram\cite{haseda} clearly suggests that the LRO is described by the order parameter $\langle S_x \rangle$. It is noted that even a slight tilt of the $c$-axis yielded asymmetry of the scattering intensity with respect to $H_{\rm c}$\cite{toda3}. Figure \ref{f5} shows the dependence of integrated intensity on temperature for the neutron magnetic reflection at 7.0, 7.5, 8.0 and 8.5 T ($H {\small /\!\!/} c$). 
Critical point exponents $\beta$ for the condensation growth have been roughly estimated by fitting to the following equation,
\begin{equation}
\sqrt{I} \propto (T_{\rm N}-T)^{\beta}.
\end{equation}
 The values of $\beta$ varied with field, each being 0.30 (7.0 T), 0.22 (7.5 T), 0.27 (8.0 T) and 0.33 (8.5 T). It should be noted, that the value of $\beta$ at 8.5 T (0.33) is near to the value of $\beta$ (0.35) for a Bose Einstein Condensation of the 3D-XY system, in spite of the quasi-one dimensional interaction of the system.
 
 To further investigate the ordered state, we try to estimate the condensation fraction of magnon. Actually, the magnetic structure of the ordered phase has not been determined yet exactly. From comparison with the isomorphous compound RbFeCl$_3$\cite{wada}, we assume a cone structure model for the magnetic structure(120$^{\circ}$ periodicity along the $a$-axis). Denoting $M_{\perp}$ and $M_{\tiny /\!\!/}$ as the antiferromagnetic and ferromagnetic ordered moment respectively, the magnetic structure factors from each component are described as
\begin{eqnarray}
|F_{hkl^{\pm}}|^2&=&\frac{1}{4}p_0^2f^2 M^2_{\perp}(1+\cos{\phi}^2)|G_{hkl}|^2, \\
|F_{hkl}|^2&=&p_0^2f^2 M^2_{\mbox{\tiny $/\!\!/$}}\sin{\phi}^2|G_{hkl}|^2, \label{eq}
\end{eqnarray}  
where $p_0$ is $0.269\times 10^{-12}$cm, $f$ is the form factor, $|G_{hkl}|^2=|\sum_{i}e^{i \bm{q}_{hkl}\cdot \bm{r}_i}|^2$ is the geometrical structure amplitudes of magnetic ions for the $(h k l)$ reflection, and $\phi$ is the angle between the scattering vector and the $c$-axis. The notation of $|F_{hkl^{\pm}}|^2$ denotes the fact that the magnetic scatterings from the $M_{\perp}$ component appear at $\vec{s}=\vec{B}_{hkl} \pm \vec{\tau}$, where $\vec{B}_{hkl}$ is the reciprocal lattice vector and $\vec{\tau}$ is the modulation vector for the magnetic structure. Because the experiments have been performed at $T \sim$ 2 K, we may approximate the Debye-Waller factor as 1, then the value of $M_{\perp}$ can be determined experimentally comparing the intensities of nuclear (1 1 0) reflection at zero field ($I_{110}$; non-magnetic) and magnetic (1/3 1/3 0) reflection ($I_{\frac{1}{3}\frac{1}{3}0}$) as following;
\begin{equation}
\frac{I_{\frac{1}{3}\frac{1}{3}0}}{I_{110}} \approx \frac{\sin^{-1}(2\theta_{\frac{1}{3}\frac{1}{3}0}) \frac{1}{4}p_0^2f^2 M^2_{\perp}|G_{110}|^2}{\sin^{-1}(2\theta_{110})|F^n_{110}|^2},
\label{magmom}
\end{equation}  
where $|F^n_{110}|^2=|\sum_{i}b_ie^{i \bm{q}_{110}\cdot \bm{r}_i}|^2$ is the square of the nuclear structure amplitude factors of all ions. For example, the value of $M_{\perp}$ is derived as 1.83 and $\langle S_x \rangle$ as 0.477 ($g^{\perp}=3.84$\cite{suzuki}) at $T=1.55$ K and $H=7.5$ T. On the other hand, the computation of $\langle S_x \rangle$ in the MFA \cite{tsuneto} gives 0.703, when the Hamiltonian parameters are chosed as $J_1$=4.4 K and $D$=17.4 K (these parameters are obtained from the fit to the experimental values of phase boundary). The dicrepancy between 0.477 (experiments) and 0.703 (MFA) is appreciable, but not too large, thus we suggest that the MFA theory adequately describes the experimental results.

 We notice here again that the obtained values of $\beta$ and $M_{\perp}$ are rather rough estimations. When the external field increases to $H \sim H_{\rm c}$, there was always a severe effect of magnetic torque which increases the tilt of the sample. Even though the obliquity was very small, the 3D-LRO state as well as condensation of magnon seemed to be affected significantly. Moreover, the data points are limited in the temperature region of $\sim 1$ K, where the precision of temperature control is $\sim 0.1$ K.
 
\begin{figure}[htbp]
\begin{center}
\includegraphics[width=8cm]{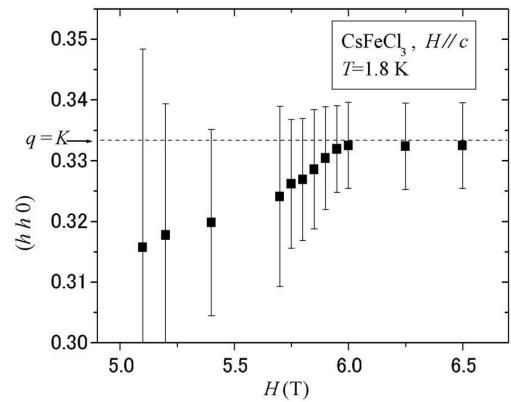}
\end{center}
\caption{The field variation of the scattering vector $\bm{q}$ and the line width along $(h h 0)$ at $T=1.8$ K in the incommensurate phase ($5<H<6$ T). Error bars denote standard deviations determined by fitting to a Gaussian curve.}
\label{f2}
\end{figure}

 So far, we have focused on the symmetrical property of $\langle S_x \rangle$ with respect to $H_{\rm c}$, in the commensurate ordered state. Meanwhile, spin fluctuation can be asymmetric with respect to $H_{\rm c}$ and such effect might appear on the incommensurate state. The origin of the incommensurate state has been attributed to the fluctuation effect by P. A. Lindg\r{a}rd\cite{lindgard} and to the dipole-dipole interaction by H. Shiba\cite{shiba}. Increasing from zero field, the magnetic reflection was first found around 5 T, at the scattering vector $\bm{q}\approx$ (0.315 0.315 0). Figure \ref{f2} shows the field variation of the scattering vector $\bm{q}$ and the line width along $(h h 0)$ at $T=1.8$ K in the incommensurate phase ($5<H<6$ T). Error bars denote the standard deviation determined by fitting to a Gaussian curve. 
The scattering vector $\bm{q}$ gradually moves to the $K$-point with increasing external field, accompanied by narrowing of the line width and increasing of the scattering intensity. Development of the coherent state and the static correlation length has been observed in the incommensurate phase. Under the field of $H > H_{\rm c}$, a satellite peak was found in coexistence with the commensurate peak at the phase boundary around 10 T at 1.8 K. Shift of the commensurate-incommensurate peak was not obvious. This result can be due to the effect of the induced ferromagnetic moment of $|S_z=+1\rangle$, which disturbs the propagation of the magnetic excitation. On the other hand, if the tilt of the $c$-axis is more than $0.5^{\circ}$, then a satellite peak should have disappeared, therefore, it indicates that the tilt of the $c$-axis would be less than $\sim 0.5^{\circ}$ at such high field. 

\section{\label{sec:level4}CONCLUSIONS}
 We have investigated the field-induced order in CsFeCl$_3$ in the whole field range up to 10 T, by means of magnetization and neutron diffraction. The field variation of the intensity of neutron magnetic reflection has clearly demonstrated that the ordered phase is described by the order parameter $\langle S_x \rangle$. The condensation growth of magnon has been investigated through the temperature dependence of $M_z$ and $M_{\perp}$, and the values of $\beta$ and the condensation fraction of magnons have been experimentally estimated. Development of the coherent state and the static correlation length has been clearly observed in the incommensurate phase at $5 < H < 6$ T. At $H > H_{\rm c}$, a satellite peak was found in coexistence with the commensurate peak at the phase boundary around 10 T, which indicates that the tilt of the $c$-axis would be less than $\sim 0.5^{\circ}$ in the whole experiments.
 
\begin{acknowledgments}
 We would like to thank K. Hirakawa for the instruction of the crystal synthesis. 
\end{acknowledgments}

\begin{references}
    \bibitem{nikuni}
    T. Nikuni, M. Oshikawa, A. Oosawa and H. Tanaka, Phys. Rev. Lett. {\bf 84}, 5868 (2000).
    \bibitem{haseda} 
    T. Haseda, N. Wada, M. Hata and K. Amaya, Physica {\bf 108}B, 841 (1981). 
    \bibitem{tsuneto}
    T. Tsuneto and T. Murao, Physica {\bf 51}, 186 (1971).
    \bibitem{yoshizawa}
    H. Yoshizawa, W. Kozukue and K. Hirakawa, J. Phys. Soc. Jpn. {\bf 49}, 144 (1980).
    \bibitem{steiner}
    M. Steiner, K. Kakurai, W. Knop, B. Dorner, R. Pynn, U. Happek, P. Day and G. McLeen, Solid State Commun. {\bf 38}, 1179 (1981).
    \bibitem{schmid}
    B. Schmid, B. Dorner, D. Petitgrand, L. P. Regnault and M. Steiner, Z. Phys. B {\bf 95}, 13 (1994).
    \bibitem{knop}
    W. Knop, M. Steiner and P. Day, J. Magn. Magn. Matter {\bf 31-34}, 1033 (1983).
    \bibitem{chiba1}
    M. Chiba, Y. Ajiro, K. Adachi and T. Morimoto, J. Phys. Soc. Jpn. {\bf 57}, 3178 (1988).
    \bibitem{toda1}
    M. Toda, T. Goto, M. Chiba, K. Adachi and N. Suzuki, J. Phys. Soc. Jpn. {\bf 68}, 2210 (1999).    
    \bibitem{toda2}
    M. Toda, T. Goto, M. Chiba and N. Suzuki, J. Phys. Soc. Jpn. {\bf 71}, 930 (2002).    
    \bibitem{suzuki}
    N. Suzuki and J. Makino, J. Phys. Soc. Jpn. {\bf 64}, 2166 (1995).
    \bibitem{hori}
    H. Hori, I. Shiozaki, M. Chiba, T. Tsuboi and M. Date, Physica B {\bf 155}, 299 (1989).    
    \bibitem{hirakawa}
    K. Hirakawa and K. Ubukoshi, Solid State Phys. {\bf 14}, 625 (1979).
    \bibitem{toda3}
    M. Toda, T. Goto, M. Chiba, S. Ueda, K. Nakajima, K. Kakurai, R. Feyerherm, A. Hoser, H. A. Graf and M. Steiner,
    J. Phys. Soc. Jpn. {\bf 70}, Suppl. A, 154 (2001).
    \bibitem{chiba2}
    M. Chiba, S. Ueda, T. Yanagimoto, M. Toda and T. Goto, Physica B {\bf 284}, Part 2, 1529 (2000).
    \bibitem{wada} 
    N. Wada, K. Sumiyoshi, T. Watanabe and K. Amaya, J. Phys. Soc. Jpn. {\bf 52}, 1893 (1983).  
    \bibitem{lindgard}
    P. A. Lindg\r{a}rd and B. Schmid, Physical Review B {\bf 48}, 13636 (1993).    
    \bibitem{shiba}
    H. Shiba and N. Suzuki, J. Phys. Soc. Jpn. {\bf 51}, 3488 (1982).   
\end{references}

\end{document}